\documentstyle[11pt,aasms4]{article}
\tighten
\slugcomment{Draft printed: ~~\today}

\begin{document}

% Needed TeX macros
\def\COBE{{\sl COBE\/}}
\def\wisk#1{\ifmmode{#1}\else{$#1$}\fi}
\def\um     {\wisk{{\rm \mu m}}}
\def\etal   {et~al.}
\def\deg    {\wisk{^\circ\ }}

\hfuzz=10pt \overfullrule=0pt
 
\pretolerance=10000
\raggedright

\pretolerance=1000	% hyphenation on

\title{
THE SPECTRUM OF THE EXTRAGALACTIC FAR INFRARED BACKGROUND FROM THE \COBE\altaffilmark{1} 
FIRAS OBSERVATIONS}

\author{ D.J. Fixsen\altaffilmark{2,3}, E. Dwek\altaffilmark{4},
J. C. Mather\altaffilmark{4}, C. L. Bennett\altaffilmark{4},
and R. A. Shafer\altaffilmark{4} }

\altaffiltext{1}{The National Aeronautics and Space Administration/Goddard 
Space Flight Center (NASA/GSFC) is responsible for the design, development, 
and operation of the {\it Cosmic Background Explorer} ({\it COBE}).
Scientific guidance is provided by the {\it COBE} Science Working Group.
GSFC is also responsible for the development of the analysis software and
for the production of the mission data sets.}
\altaffiltext{2}{Hughes STX Corporation, 
                 Code 685, NASA/GSFC, 
                 Greenbelt MD 20771.}
\altaffiltext{3}{e-mail: fixsen@stars.gsfc.nasa.gov}
\altaffiltext{4}{Laboratory for Astronomy and Solar Physics,
                 Code 685, NASA/GSFC, 
                 Greenbelt MD 20771.}

\begin{abstract}

The {\it COBE} FIRAS data contain foreground emission from interplanetary, 
Galactic interstellar dust and extragalactic background emission. We use 
three different methods to separate the various emission components, and 
derive the spectrum of the extragalactic Far InfraRed Background (FIRB). 
Each method relies on a different set of assumptions, which affect the FIRB 
spectrum in different ways. Despite this, the FIRB spectra derived by these 
different methods are remarkably similar. The average spectrum that we derive 
in the $\nu = 5 - 80~$cm$^{-1}$ (2000 -125 \um) frequency interval is: 
$I(\nu) = (1.3\pm0.4)\times10^{-5}\ (\nu /\nu_0)^{0.64\pm.12}\ 
P_{\nu}(18.5\pm 1.2~$K), where $\nu_0=100~$cm$^{-1}$ ($\lambda_0=100\ \um$) and
$P$ is the Planck function. 
The derived FIRB spectrum is consistent with the DIRBE 140 and 240 \um\  
detections. The total intensity received in the 5 - 80~cm$^{-1}$ frequency 
interval is 14 nW m$^{-2}$ sr$^{-1}$, and comprises about 20\% of the total 
intensity expected from the energy release from nucleosynthesis throughout 
the history of the universe.
\end{abstract}


\section{INTRODUCTION} 

The FIRAS (Far~Infrared~Absolute~Spectrophotometer) instrument aboard the Cosmic Background Explorer 
({\it COBE}) satellite (Boggess \etal\ 1992 and references therein; Mather, Fixsen, \& Shafer 
1993) was designed to measure the spectrum of the Cosmic Microwave Background
(CMB), the Galaxy, and the Far InfraRed Background (FIRB). The FIRAS instrument 
and calibration are discussed by Mather, Fixsen \& Shafer (1993) and 
Fixsen \etal\ (1994b). The FIRAS observations of the CMB are discussed by 
Mather \etal\ (1990, 1994), and Fixsen \etal\ (1994a, 1996, 1997a). The FIRAS 
observations of the Galaxy are discussed in Wright \etal\ (1991), Bennett 
\etal\ (1994), and Reach \etal\ (1995). Puget \etal\ (1996) used the Pass 3
FIRAS observations to make a tentative background determination simular to
one of the methods used here.

At frequencies below $\sim$ 20 cm$^{-1}$ (500 \um), the FIRB is overwhelmed 
by the CMB. However, the CMB can be subtracted from the data since its emission 
is spatially uniform, and it has a Planck spectrum. The dipole anisotropy
of the CMB must be included
but the other anisotropy is not large enough to be a problem. At frequencies 
above $\sim$ 100 cm$^{-1}$ (100 \um),beyond the FIRAS range, the observed 
spectrum is dominated by the zodiacal 
emission, which is much more complex, since its intensity and spectrum are 
spatially varying. In between these frequencies, the observed spectrum is 
dominated by Galactic emission, which can be identified by its distinct
(from zodiacal) spectrum, and its spatial variation over the sky. After the 
subtraction of these foreground emissions, any isotropic residual emission
may still contain a uniform component from the
solar system or the Galaxy, which must be estimated and separated from the 
FIRB. In this paper we examine three methods, which give a consistent result
for the FIRB.

\section{THE OBSERVATIONS}

The FIRAS Pass 4 data consist of spectra between 2 and 96 cm$^{-1}$ 
(5000 to 104 \um\ wavelength) in each of 6063 pixels on the sky (there 
are 6144 pixels in the full sky) with 210 frequency bins per spectrum. 
They were calibrated using the method 
described by Fixsen \etal\ (1994b), with the improvements noted by Fixsen 
\etal\ (1996). The Pass 4 FIRAS data have lower noise and lower systematic 
errors than the previous FIRAS data releases (FIRAS Explanatory Supplement). 
A weighted average of all of the FIRAS data was used for this analysis.
Random uncertainties are $\sim0.05$ MJy/sr ($\nu<50~$cm$^{-1}$) and up to
2 MJy/sr at $\nu=90$ for 10\% of the sky. Systematic uncertainties are
approximately the same magnitude. The small instrumental offsets deduced
from the cold external calibrator are what allow this absolute measurement.
The instrumental uncertainties are small
relative to the uncertainties associated with the foreground removal. 
We use data from the Diffuse InfraRed Background Experiment (DIRBE) (DIRBE 
Explanatory Supplement) to sort and analyze the FIRAS data, 
and to subtract the DIRBE$-$determined zodiacal dust emission (Kelsall \etal\ 
1998) extrapolated to the FIRAS frequencies, from the FIRAS data. 

\section{SEPARATION OF THE GALACTIC EMISSION COMPONENT}

The infrared (IR) emission from interplanetary dust particles is a minor 
contributor to the FIRAS emission and then only in the higher ($>$ 70 cm$^{-1}$) 
frequency channels. This emission component is removed from the 
data by using the model developed from the analysis of the DIRBE data 
(Kelsall \etal\ 1998). 
The major challenge facing the search for the FIRB in the FIRAS data is the 
removal of Galactic emission. 

We have used three different and complementary 
methods for separating the Galactic emission from the FIRB. The first relies 
on the spatial variability of the Galactic emission component. Assuming a 
spatially invariant spectral shape (i.e. color) for this component, we use 
its spatially varying intensity to distinguish it from the isotropic FIRB. The 
second method uses H~I 21 cm and [C~II] (157.7 \um) line emission, which trace 
the atomic and ionized gas components of the interstellar medium (ISM), as 
templates for the Galactic emission component. The third method uses the 
DIRBE 140 and 240 \um\ detections of the FIRB (Hauser \etal\ 1998) to define 
templates of Galactic emission. 

Each of these methods relies on some assumptions, and therefore has 
its shortcomings. However, the assumptions underlying each method are 
different, so the similarity between the derived FIRB spectra 
suggests that this spectrum represents a robust estimate of the FIRB.

\subsection{A FIRAS Color Template}

The FIRAS sky spectra, $S(\ell,b;\nu)$, are a function of frequency $\nu$ and 
position \{$\ell$, $b$\}, where $\ell$ and $b$ are Galactic coordinates. 
Since the data are binned into discrete pixels, $p$, and frequencies, $\nu$, 
we can also write the FIRAS intensity in each pixel as $S_{p\nu}$. 
As the first step in the analysis we sort the sky into 10 bins. The first bin 
is the 10\% of the sky which is darkest according to the DIRBE 100 \um\ map,
which has had the zodiacal emission removed with the DIRBE zodiacal model.
The next bin is the next 10\%, etc. 
Since the DIRBE data are independent of the FIRAS data we have introduced no 
noise biases by this sorting. Since the amount of FIRAS calibration
data is about 10\% of the sky data one can think of them as a zeroth bin.
Because the calibration uncertainties of the FIRAS data are of the same 
magnitude as the random uncertainties in 10\% of the sky, there is
little to be gained by averaging more than 10\% of the sky.

We form the average intensity in each bin $k$, $S_{k\nu}~~(k=1,2...10)$,
using the pixel weights of the FIRAS data. A map of the sky bins
is shown in Fig 1. The sorting is primarily based on the intensity of 
Galactic emission. The resulting spectra are shown in Fig 2a. Several items 
are apparent from the figure: (1) the CMB radiation dominates 
below 20 cm$^{-1}$; (2) zodiacal emission is evident above $\sim$ 70 cm$^{-1}$, 
especially in the lower intensity bins; (3) between these frequencies the
emission is dominated by the Galaxy, which has a similar spectrum in all bins,
particularly 
in the dimmer bins; (4) the [C~II] (63 cm$^{-1}$; 158 \um) line is apparent in 
all spectra; and (5) the [N~II] line (48 cm$^{-1}$; 205 \um) line is visible 
in the brighter bins and the brighter bins have a higher proportion of high 
frequency radiation (they are warmer). Although noise is evident in the dimmer 
bins, the signal to noise ratio in the 20 to 70 cm$^{-1}$ frequency interval 
is good for all bins.

Figure 2b shows the spectra of the bins after the removal of the CMB and the 
zodiacal dust emission. The CMB model is made from the low frequency FIRAS 
data, while the zodiacal model is determined from the DIRBE data 
(see Kelsall \etal\ 1998). The zodiacal dust model was interpolated and 
extrapolated from the DIRBE 240 \um\ and 140 \um\ zodiacal models
to the FIRAS frequencies, except that for frequencies below 
40 cm$^{-1}$ the slope was steepened by multiplying by a factor $(\nu-20)/20$
for frequencies between 20 and 40 cm$^{-1}$. Below 20 cm$^{-1}$ no zodiacal
correction was made. There is no evidence of zodiacal 
emission below 20 cm$^{-1}$ in any analysis, so the zodiacal spectrum
must steepen between 20 cm$^{-1}$ and 60 cm$^{-1}$.

The CMB model must be good to more than 3 orders of magnitude for the low 
frequency Galactic spectra to emerge. The effect of the CMB dipole (also shown 
in Fig 2a) must also be included in the CMB model. In contrast the zodiacal 
emission contributes only a small fraction to the emission at frequencies below 
80~cm$^{-1}$, and we can tolerate 20\% errors in its estimate.

The spectrum of bin 1 (the faintest 10\% of the sky) provides a very robust
upper limit on the FIRB. The signal to noise ratio is high and the systematic
errors are small relative to the signal. But this is an upper limit, not
a detection.

We model the average intensity in each bin as a sum of two components, 
a uniform component, $U$, and a spatially varying Galactic component, $G$, 
that may be spectrally distinct from $U$. The average intensity in each bin 
is then

\begin{equation}
S_{k\nu} = U_\nu + g_k G_\nu,
\end{equation}

\noindent
where $U_\nu$ is a uniform spectrum containing the FIRB, $G_\nu$ is the 
Galactic spectrum and $g_k$ is a parameter that determines its intensity in 
the $k$th bin. There are two degrees of freedom that are 
unconstrained in this fit. For a given solution $(U', g', G')$, there exists a 
family of solutions: $(U'-xG',(g'+x)/y,yG')$ that provide identical fits to 
\{$S_{k\nu}$\}. The two degrees of freedom are represented by 
an additive component to the uniform spectrum, and a scaling
parameter for the Galactic spectrum. Two additional constraints provide
a unique solution to eq. (1). First, the background must be positive, 
so one solution, $U''$ is chosen with $x=x_\circ$ so that $U''=U'-x_\circ G'$ 
is at most 2 $\sigma$ negative at any frequency. Second, we scale the Galactic 
spectrum in terms of its value in the dimmest bin ($k$=1), $G''= S_1 - U''$. 
With this choice, the family of solutions is expressed as
\begin{equation}
U = U'' + \gamma G''.
\end{equation}
\noindent
The parameter $\gamma <1$, otherwise the background $U$ is brighter than 
the intensity $S_1$ in the dimmest 10\% of the sky. Furthermore, $\gamma > 0$, 
otherwise $U$ falls below the 2$\sigma$ negative value. We estimate a 
value of $\gamma\approx 0.2$. A lower value is unlikely because it (weakly) implies a negative 
background at some frequencies. A higher value unnecessarily ascribes to the 
background, radiation which is fit equally well with the known Galactic 
foreground. Also if $\gamma$ is larger, the Galactic [C~II] line is clearly 
visible in the background spectrum. Nominally the uncertainty in $\gamma$ would 
be 0.1 but in view of the systematic nature of the possible errors a $\sigma$ 
of 0.2 is appropriate.

Should there be a systematic variation of temperature with intensity it would
couple with the assumption of a single color for the Galactic emission leading 
to errors in the calculated background shape and intensity. Part or all of the 
``observed" background could be due to this effect for this method.

When small bins are used there is an indication that the temperature is 
variable, and the temperature depends on Galactic longitude. The 
spectrum of the Galactic plane is different from that of the high latitude
Galaxy, but using the faintest 20 to 80\% of the sky gives the same results to 
a good approximation. The decomposition for the faintest 50\% of the sky is 
shown in Figure 3a. The noise of the FIRAS data increases with frequency and
at frequencies $\nu>80$~cm$^{-1}$ may be affected by zodiacal emission. The 
feature at $\sim20$~cm$^{-1}$ is in the low weight data between the low and 
high frequency channels of the FIRAS and is not significant.

\subsection{An H~I and [C~II] Line Emission Template}

The Galactic foreground emission is estimated and removed by subtracting 
the fraction of the observed emission that correlates with templates of the 
Galactic IR emission. The H~I 21 cm and [C~II] 158 \um\ lines are good candidates for such
templates because they are definitely of Galactic origin, they are both 
strong, and nearly fully sampled across the sky. While they may not correspond
directly to the neutral and ionized gas phases, they are not perfectly 
correlated with each other and if they are differently associated with the 
different ISM phases then linear combinations of them could be used to 
trace the emission of the neutral and ionized gas, the major
components that give rise to the high latitude Galactic FIR emission 
(Reach \etal\ 1995). 

The H~I map was used by Puget \etal\ (1996). We use the 
AT\&T Bell Laboratories H~I survey map from Stark \etal\ (1992) convolved with 
the FIRAS beam. The [C~II] map is from the FIRAS so it already has the right 
beam profile and sky coverage. We have also tested an [N~II] 205 \um\ template,
from the FIRAS data, but at high latitude the [N~II] emission is small and 
there is no significant correlation. For both the H~I and the [C~II] maps
the zero point is not a significant uncertainty.  The relation between the
line emission and the continuum emission is a far larger uncertainty.

The relation between H~I and the 
IR emission is clearly not linear. At moderate H~I column densities the 
molecular mass fraction of a cloud becomes significant, and IR emission from 
the molecular gas causes an excess of IR emission over that expected from the
H~I-IR correlation plot (Boulanger \etal\ 1996, Spaans \& Neufeld 1997). We
therefore include a quadratic term in H~I in the fit, as well as the 
[C~II] emission as a second tracer. Both line maps are renormalized to an 
average value of 1 over the faintest half of the sky.
The intensity in a given sky pixel is: 
\begin{equation}
S_{p\nu} = U_\nu +h_p H_\nu+h_p^2 G_\nu+c_p C_\nu
\end{equation}
\noindent
where $p$ is the sky pixel index, $h_p$ is the H~I map, $c_p$ is the [C~II] 
map and $U, G, H, C$ are spectra determined from the fit (the $U_\nu$
is again the background but not necessarily that of section 3.1). The linear 
term is small indicating that the quadratic term absorbs most of the 
power. Even though the linear term is negative at some frequencies the 
combination of linear plus quadratic is positive.

We must eliminate the brightest 2/3 of the sky and the south polar region as we 
have no H~I data with well accurate sidelobe corrections there. These limit the 
fit to $\sim25\%$ of the sky in order to get a stable fit. The fit to the [C~II] 
map is quite stable over a wide range of cutoffs but the $H$ spectrum fit, even 
with the inclusion of the quadratic term, is only stable over the dimmest 
parts of the sky. The results of the fit are shown in figure 3b.

Since the relation between the H~I and the Galactic emission is not linear
there is a concern that a quadratic is not the right answer either. Forcing
the fit to be linear and cutting enough data to make it fit leaves a very short
lever arm for the extrapolation. The [C~II] picks up other radiation but there
still might be more Galactic radiation not traced by either the H~I or [C~II].

The expected noise of the fit spectra is evident. This is exaggerated by
the high correlation between the $h_p$ and $h_p^2$. Still there is a clear 
signal at lower frequencies.

\subsection{A DIRBE 140 and 240 \um\ Emission Template}

In many ways the easiest approach is to use the DIRBE maps and extragalactic 
background detections at 140 and 240 \um\ to determine the FIRB from the FIRAS 
data. Adopting the DIRBE FIRB backgrounds (Hauser \etal\ 1998) at 140 and 
240 \um\ of 1.17 and 1.12 MJy/sr respectively, we construct Galactic emission 
templates by subtracting these backgrounds from the zodiacal light subtracted 
DIRBE maps convolved with the FIRAS beam. This combination 
has already been shown to model the Galaxy quite well over a large part of 
the sky (Fixsen \etal\ 1997a).
We represent the pixel intensity at each FIRAS frequency $\nu$ in terms of a linear combination of the two DIRBE
templates as follows:

\begin{equation}
S_{p\nu} = U_\nu +(D^{240}_p-D^{240}_\circ) G^{240}_\nu + 
(D^{140}_p-D^{140}_\circ) G^{140}_\nu
\end{equation}
\noindent
where $D^{\lambda}_\circ$ and $D^{\lambda}_p$ are, the DIRBE background and 
map intensity at FIRAS pixel $p$, at wavelength $\lambda$. By correlating the 
FIRAS map to both the 140 and 240 \um\ DIRBE templates, we allow for temperature 
variations in the Galactic foreground. The background is obtained by 
extrapolating the correlations with the DIRBE templates to the DIRBE background 
values of $D^{\lambda}_\circ$. Figure 3c presents the results of the fit. 
To be consistent with the other fits, we plot the results for the dimmest half 
of the sky. However, the result is very robust and essentially identical 
results can be obtained for any
region of the sky that excludes the Galactic plane.

Of course the uncertainties in the DIRBE FIRB determination at 140 and 240 \um\ 
propagate the determination of the FIRB here. We have shown (Fixsen 1997b)
that the FIRAS and DIRBE calibrations are consistent, so it
is not surprising that the two data sets give consistent results when
the amplitude of the background determination for the DIRBE is assumed.
The DIRBE maps are only used to determine the {\it Galaxy} model; 
the isotropic component need not have been consistent with the FIRAS 
determination so this method is partially independent.

\section{DISCUSSION}

The three methods just described have different assumptions, weaknesses and
strengths.
The first method assumes that the Galactic spectrum is fixed, and that only its 
intensity varies with position. Temperature variations are observed to occur on 
various scales (Reach et al. 1995; FIRAS Exp Sup 1997; Lagache et al. 1998). 
Based on the DIRBE analysis, we expect these variations to have a 
only a small effect on our results.
The DIRBE high latitude data at 240 \um\ were analyzed using both a single-
and a variable-spectrum model to describe the Galactic foreground emission
(Arendt et al. 1998), with no significant difference in the derived background. 
  
The second method does not assume a fixed Galactic spectrum, but almost all of 
the power in the fit is in a single component indicating that there is little
systematic correlation of temperature with intensity of the H~I 21 cm line. 
This method
solves directly for the dust emission from the dominant neutral and ionized 
phases of the ISM. It therefore avoids a potential problem of the third method 
(which relies on the DIRBE analysis) using backgrounds derived from the 
IR-to-H~I correlation only. Two problems with this method are, that at the 
FIRAS resolution, the correlation between the IR and the H~I line intensity 
is non-linear and that it ignores the molecular regions. The first was treated
by using a quadratic fit to the relation. This choice is not unique, but it 
is supported by observations of the IR to H~I correlation in the Galaxy 
(Dall'Oglio et al. 1985; Reach, Koo, \& Heiles 1994). Molecular clouds
are less abundant at the high Galactic latitudes that are used here.

The third method relies on the accuracy of the DIRBE determination of the FIRB.
The main uncertainty in the background determination are those associated with
the determination of the foreground emission, and they are discussed in detail
by Hauser et al. (1998). The use of a DIRBE template for the FIRAS background
determination introduces another uncertainty, namely the consistency between the
FIRAS and DIRBE calibration. Fixsen et al. (1997b) show that the most significant
difference in the calibration is in the 240 \um\ DIRBE band, but using the
FIRAS instead of the DIRBE calibration for that band introduces a very small
change in  the background, from a value of 13.6 to 12.7 nW m$^{-2}$ sr$^{-1}$
(Hauser et al. 1998).

Overall, the three methods yield a consistent spectrum for the FIRB, an 
encouraging result considering the substantial differences in the three 
approaches (see fig 4a). The weaknesses of each method are compensated by the
other methods. The color method assumes a single spectrum, but the other methods
allow for color variation in the Galactic foreground. The line emission method 
uses a quadratic fit over 25\% of the sky, but the other methods use linear
fits which are insensitive to the fraction of sky.  The DIRBE method only uses
the correlation between H~I and the foreground in small regions, but the 
line emission method includes ionized gas and the color method makes no assumptions
about the gas.

In all three FIRB spectra the background peaks at $\sim 50~$cm$^{-1}$ 
($\sim 200$ \um), and exhibits a definite turnover at the 
higher frequencies. The higher frequencies are more affected by both noise and 
systematic effects. The average of the three spectra can be fit by:

\begin{equation}
I_\nu=(1.3\pm 0.4)\times 10^{-5}(\nu/\nu_\circ)^{.64\pm.12}\ 
P_\nu(18.5\pm1.2~K), 
\end{equation}
in the 5 to 80 cm$^{-1}$ frequency range ($\lambda$ between 125 and 2000 \um) 
where $\nu_\circ=100~$cm$^{-1}$, and $P_\nu$ is the familiar Planck
function. The uncertainties are highly correlated, with correlations of 
.98 for the intensity and index, $-$.99 for the intensity and temperature. and
$-$.95 for the index and temperature.

Figure 4b compares the analytical fit to the FIRB derived here, to the 
tentative background derived by Puget et al. (1996). The spectra differ 
significantly at $\nu \sim 35 - 60~$cm$^{-1}$, probably the result of the 
difference in subtraction of dust emission related to H$^+$.
The crosses in the Figure represent the nominal DIRBE detections at 140 and 240 
\um, while the diamonds represent the DIRBE detections using the FIRAS 
calibration. While the effect of the FIRAS calibration is larger at 140 \um, 
the uncertainty in the calibration is larger at this wavelength. 
So, the FIRB derived here is consistent with that derived by the 
DIRBE, within the uncertainty of the DIRBE$-$FIRAS calibration. The range of
values for the FIRB at 100 \um\ represent the upper limit derived by 
Kashlinsky, Mather, \& Odenwald (1996) from a fluctuation analysis of the 
100 \um\ DIRBE maps and the lower limit is derived by Dwek et al. (1998) from 
the 140 and 240 \um\ DIRBE detections.

Dwek \etal\ (1998) show that the uniform DIRBE residuals cannot be produced by 
any local (solar system or Galactic) emission sources. Hence, the uniform residual 
derived here is most likely of extragalactic origin. The total flux received 
in this wavelength region is 14 nW m$^{-2}$ sr$^{-1}$, or about 20\% of the 
total expected flux of about 70 nW m$^{-2}$sr$^{-1}$ associated with the 
production of metals throughout the history of the universe in some 
models (Dwek \etal\ 1998).

\acknowledgements
We thank the many people involved in processing the FIRAS data.
We thank the DIRBE team for helpful discussions and the zodiacal model, 
G. Hinshaw for help in producing the plots, and the referee, W Reach for
his helpful comments.
This work was supported by the Office of Space Sciences at NASA Headquarters.

\clearpage
\typeout{FIGURE CAPTIONS}
\begin{figure}
\caption{Ten regions of the sky as defined by the DIRBE 100 \um\ intensities.
The brightest region is white. The black streak and spots are pixels not used
in this analysis.}
\label{regions}
\end{figure}

\begin{figure}
\vspace {.8in}
\caption{a)Average spectra for the ten regions shown in Figure 1. b)The spectra 
after the removal of the CMB and the zodiacal emission models.
The CMB and CMB dipole are shown with dotted lines and the average zodiacal 
emission is shown with a dashed line.}
\end{figure}

\begin{figure}
\caption{a) Galaxy (dotted) and minimum background (solid) spectra determined 
from method 1 (\S 3.1). b)The background $U$ and the spectra $C$, $H$, and 
$G$ derived by using the H~I and [C~II] line emission maps to
fit the FIRAS data [eq. (3)]. The line maps have been normalized to 1 over 
the dimmest half of the sky. The background is given by a solid line, 
the dotted line is $C$, the dashed one is the $H$, and the dash-dotted line
is $G$ (see \S 2.2 for more details).
c) The background derived by using DIRBE 140 and 240 \um\ Galactic templates. 
The solid line is the background, and the dashed and dotted lines are the 
spectra associated with the 240 and 140 \um\ Galactic templates. Although the 
140 \um\ spectrum is negative there is enough 240 \um\ emission to make 
the Galaxy model positive everywhere.}
\label{dirbeSpectra}
\end{figure}

\begin{figure}
\caption{a) A comparison of the background spectra derived by the three 
different methods. The solid line is the color model ($\S$ 3.1), the dotted 
line is the H~I-[C~II] model ($\S$ 3.2), and dashed the line is the DIRBE 
model ($\S$ 3.3). The smooth curve
is $1.3\times 10^{-5}(\nu/\nu_\circ)^{.64} P_\nu(18.5~$K),
where $\nu_\circ=100~$cm$^{-1}$ and $P$ is the Planck function.
b)A comparison of the average background spectrum derived in this paper 
(given by its analytical representation, solid line) with the DIRBE 
determination (crosses) at 140 and 240 \um\ (Hauser \etal\ 1998), and the 
range of allowable 100 \um\ intensities (Dwek \etal\ 1998). The diamonds are 
the DIRBE determinations recalibrated with the cross calibration in 
Fixsen (1997b). The light lines represent 1 $\sigma$ errors on the derivation.
Also shown is the tentative determination of the background by 
Puget \etal\ (1996) with (dashed line) and without (dash triple dot line)
their H$^+$ correction.}
\end{figure}
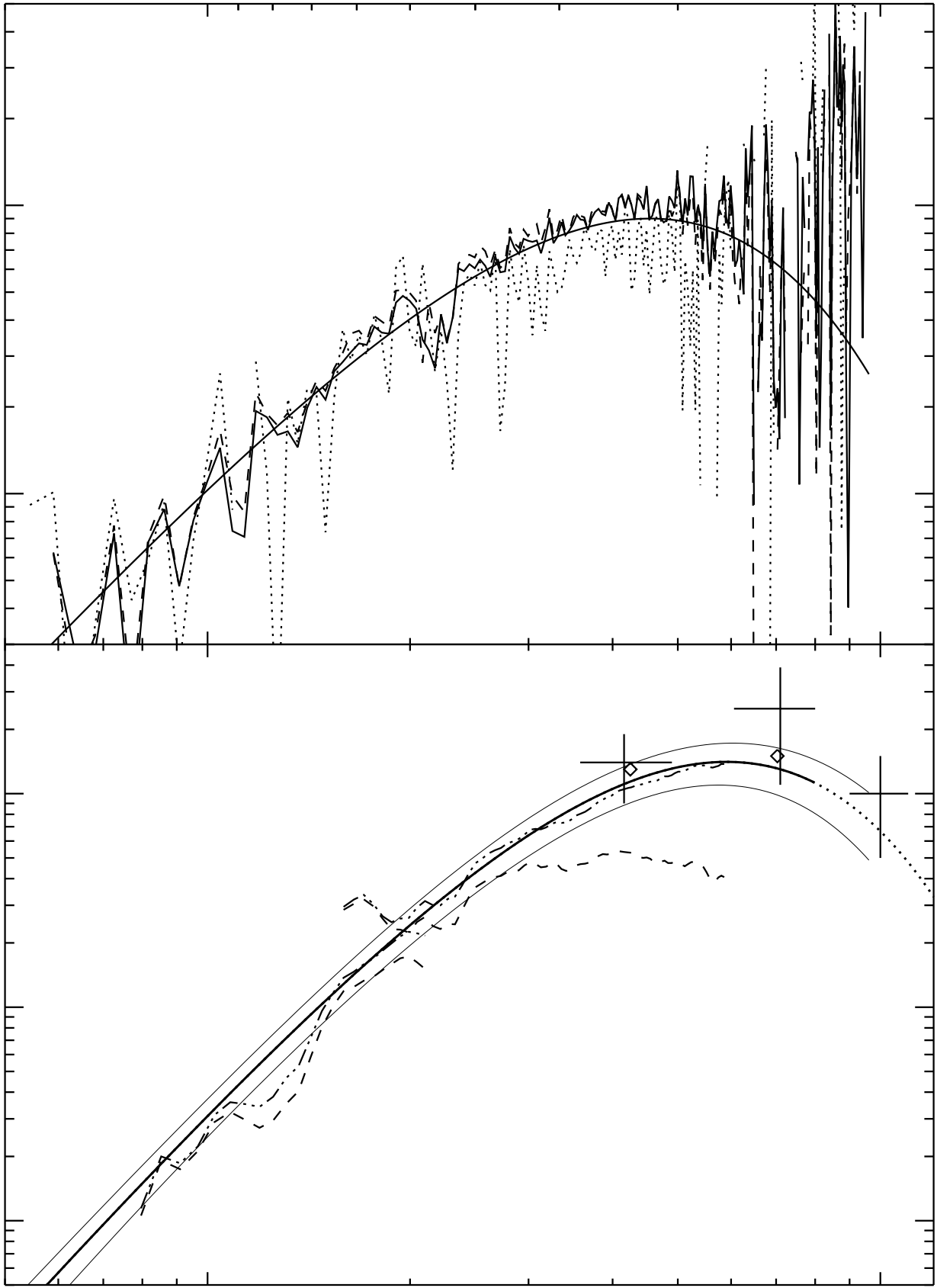
\end{document}